\documentclass[aps,prl,superscriptaddress,10pt,twocolumn,amsmath,floatfix]{revtex4-1}

\usepackage{graphicx}
\usepackage{amsmath, amsthm, amssymb}
\usepackage{url}
\graphicspath{{figs/}}
\begin{document}
\title{Patterns and Stability of Coupled Multi-Stable Nonlinear Oscillators}
\author{G. Bel}
\affiliation{Department of Solar Energy and Environmental Physics, Blaustein Institutes for Desert Research and Department of Physics, Ben-Gurion University of the Negev, Sede Boqer Campus 84990, Israel}
\affiliation{Center for Nonlinear Studies (CNLS), Los Alamos National Laboratory, Los Alamos, NM 87545, USA}
\author{B. S. Alexandrov}
\author{A. R. Bishop}
\author{K. \O. Rasmussen}
\affiliation{Theoretical Division, Los Alamos National Laboratory, Los Alamos, NM 87545, USA}
\date{\today}
\begin{abstract}
Nonlinear isolated and coupled oscillators are extensively studied as prototypical nonlinear dynamics models. Much attention has been devoted to oscillator synchronization or the lack thereof. Here, we study the synchronization and stability of coupled driven-damped Helmholtz-Duffing oscillators in bi-stability regimes. We find that despite the fact that the system parameters and the driving force are identical, the stability of the two states to spatially non-uniform perturbations is very different. Moreover, the final stable states, resulting from these spatial perturbations, are not solely dictated by the wavelength of the perturbing mode and take different spatial configurations in terms of the coupled oscillator phases.
\end{abstract}

\maketitle
\section{Introduction}
\label{sec:intro}

At present, our understanding of low-dimensional dynamical systems is very advanced \cite{D_systems}, and much of the research focus has therefore shifted to dynamical systems with many degrees of freedom, in which the study of interesting cooperative behavior and pattern formation has intrigued researchers \cite{coop}. As a result, the past decades have seen several sustained bursts in research activities focused on such topics as discrete nonlinear breathers \cite{Flach1998}, chimera solutions \cite{Abrams}, etc. The coherent and self-organized motion of coupled nonlinear oscillators is the paradigm underpinning most of these phenomena. Different synchronization regimes, defined by various degrees of temporal locking between the variables and/or the measurable properties that characterize oscillations, constitute a diverse variety of collective behavior that all emerge from the combination of individual dynamics and interactions. 

Initial models of self-organized dynamics in ensembles of coupled oscillators were inspired by the observation of synchronization phenomena in biological populations. The quantitative representation of an oscillating element in terms of a phase measured along its cyclic trajectory, which underlies Kuramoto's celebrated phase-oscillator model \cite{kuramoto1975self,kuramoto1984chemical}, has been applied in numerous variations to the description of a broad class of chemical and biological systems, ranging from catalytic surface reactions to neural networks and ecosystems. Complex spatial patterns are most often observed in systems driven out of equilibrium. Typically, the patterns emerge when relatively simple systems are driven into unstable states that will deform dramatically in response to small perturbations. As the patterns arise from an instability, the pattern-forming behavior is likely to be extremely sensitive to small changes in the system parameters \cite{Vanossi2000,Xu2014,Palmero2016,Gao2018}. 

Following earlier work \cite{Bel2018double}, we are particularly interested in exploring the interplay between harmonic and subharmonic extended states. For example, as is well understood, period doubling occurs through a bifurcation in a dynamical system in which a slight change in a parameter value in the system’s equations leads to switching to a new behavior with twice the period of the original system \cite{Ulrich1985,Kalmar-nagy2011}.
A  periodically  driven  nonlinear  oscillator  typically responds with solutions that oscillate at the same frequency as the periodic drive. In such systems, a period-doubling bifurcation is expressed through these solutions losing stability as solutions oscillating at half the driving frequency gain stability.

The simplest driven and damped dynamics exhibiting transitions between harmonic and subharmonic states is the Helmholtz-Duffing model. This model was used to describe various systems including logic devices \cite{Yao2012,Yao2013} and neural networks \cite{Grossberg1988,Hanggi2002}. Various characteristics and emerging phenomena were investigated in the context of this model, including effects of noise \cite{Rajasekar1998,Perkins2017}, chaos and the route to chaos \cite{Kenfack2003,Musielak2005}, and synchronization \cite{Vincent2008,Zanette2016}.

Here we use this minimal model to elucidate the phenomena of the stability and instability of the synchronized oscillation of coupled nonlinear oscillators in the presence of spatially non-uniform perturbations, as well as the transitions between harmonic and subharmonic states.   
We develop numerical approaches to follow the stability of these synchronized states, as well as their spatial non-uniform behavior as they transition away from stability.

\section{Floquet analysis of stability to non-uniform perturbations}

\begin{figure}[!h]
\includegraphics[width=\linewidth]{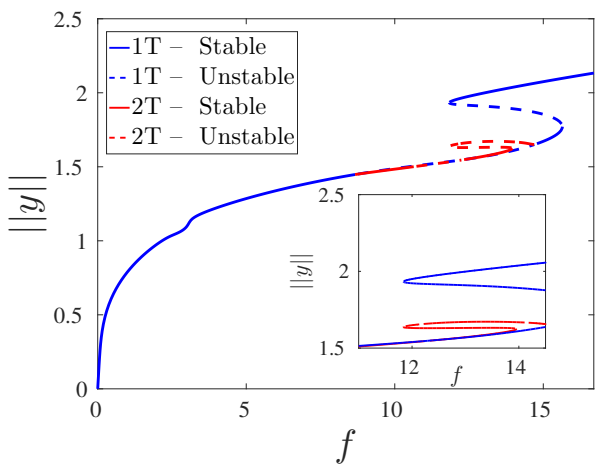}
\caption{\label{fig:L2Bif} A partial bifurcation diagram of a single Duffing-Helmholtz oscillator. The y-axis represents the L2 norm (the average was taken over one period of the oscillatory solution) of the solution, $||y||$, and the x-axis represents the amplitude of the driving force, $f$. The inset shows the bistability region. The solid lines correspond to stable solutions while the dashed lines correspond to unstable solutions. The blue (red) lines correspond to solutions with the same (double) period of the driving force (denoted as $1T$ and $2T$, respectively). The bifurcation diagram was numerically derived using the AUTO numerical continuation program \cite{doedel1981auto}.}
\end{figure}

The dynamics of the coupled asymmetric Duffing-Helmholtz oscillators is described by the following equation:

\begin{equation}\label{eq:cdh}
\frac{d^{2}y_{n}}{dt^{2}}=-y_{n}-y_{n}^{3}-by_{n}^{2}-\gamma \frac{dy_{n}}{dt%
}+f\cos \left( \omega_0 t\right) +c\Delta y_{n}, 
\end{equation}

where $\Delta y_{n}\equiv\left(y_{n+1}(t)+y_{n-1}(t)-2y_n(t)\right)$.
$b$ is the coefficient describing the asymmetry of the potential, $\gamma$ is the friction coefficient, $f$ and $\omega_0$ are the amplitude and frequency of the driving force, and $c$ is the coupling strength between nearest neighbor oscillators. We assume periodic boundary conditions (although our numerical studies revealed that for chains of 11 oscillators (or more), the results are not sensitive to the boundary conditions; we tried reflecting and free boundary conditions). Throughout this paper, where numerical results are presented, we used the following parameters unless otherwise specified: $b=0.5$, $\gamma=0.2$, and $\omega_0=1$. Within the bistability region, we used $f=13.3$ and $c=0.1$.
In Fig. \ref{fig:L2Bif}, we present the bifurcation diagram for a single oscillator versus the amplitude of the driving force. For low amplitudes of the driving force, the solution has the period of the driving force, a solution that we denote as $1T$. For larger amplitudes of the driving force, this solution loses stability, and the stable solution has a period that is twice the period of the driving force, a solution that we denote as $2T$.

In order to study the stability of a chain of coupled oscillators, we decompose the solution as follows:

\begin{equation}\label{eq:pert}
y_{n}=y_{0}+z_{n}, 
\end{equation}

where $y_{0}$ is the solution in the absence of coupling ($c=0$) and $z_{n}$ is the deviation from that solution. Assuming that the deviation from the uniform solution is small, one may derive the linearized equations describing the dynamics of the perturbations (the deviations from the uncoupled solution)

\begin{align}
\frac{d^{2}z_{n}}{dt^{2}} =&-\left(1+3y_{0}^{2}(t)+2by_{0}(t)\right)z_{n}-\gamma 
\frac{dz_{n}}{dt}+c\Delta z_{n}.
\end{align}

Assuming periodic boundary conditions, one can expand the perturbation in a Fourier series as:

\begin{equation}\label{eq:pertexp}
z_{n}=\displaystyle\sum\limits_{k=1}^{N}\zeta_{k}e^{i2\pi kn/N}\left( t\right).
\end{equation}

Due to the linear nature of the equations, one obtains for the Fourier
amplitudes, $\zeta_{k}$, the following uncoupled equations:

\begin{align}\label{eq:pertfdyn}
\frac{d^{2}\zeta_{k}\left( t\right) }{dt^{2}} =&-\zeta_{k}\left( t\right)\left(1+3y_{0}^{2}(t)+2by_{0}(t)\right) -\gamma
\frac{d\zeta_{k}\left( t\right) }{dt}\nonumber \\
&+2c\left( \cos \left( 2\pi k/N\right) -1\right) \zeta_{k}\left( t\right). 
\end{align}

The equations are linear but include parametric time dependence through the
dependence on the periodic uniform solution, $y_0(t)$ (which depends on the model parameters and, in multi-stability regions, also on the initial conditions). In order to analyze the stability of the spatially uniform solution, we determine the Floquet multipliers. 
For convenience, we rewrite the equation above as a set of two
coupled first-order equations:

\begin{align}\label{eq:pertfset}
\frac{d\zeta_{k}\left( t\right) }{dt}=&\dot{\zeta}_{k}\left( t\right)  \\
\frac{d\dot{\zeta}_{k}\left( t\right)}{dt}=&-a\left( t\right) \zeta_{k}\left(
t\right) -\gamma \dot{\zeta}_{k}\left( t\right),\nonumber
\end{align}

where

\begin{align}\label{eq:pertfseta}
a\left( t\right)=& 1+4c\sin ^{2}\left( \frac{\pi k}{N}\right) +\left(
3y_{0}(t)+2b\right) y_{0}(t).
\end{align}

In matrix form, we can write it as

\begin{align}\label{eq:matpertfdyn}
\frac{d}{dt}\left( 
\begin{array}{c}
\zeta_{k}\left( t\right)  \\ 
\dot{\zeta}_{k}\left( t\right) 
\end{array}%
\right) =A\left( t\right) \left( 
\begin{array}{c}
\zeta_{k}\left( t\right)  \\ 
\dot{\zeta}_{k}\left( t\right) 
\end{array}%
\right) ,
\end{align}
where%
\begin{align}
A\left( t\right) =\left( 
\begin{array}{cc}
0 & 1 \\ 
-a(t) & -\gamma 
\end{array}%
\right).
\end{align}

It is important to note that $a(t)$ is a periodic function whose period is set by the uniform solution (which is equivalent to the solution for uncoupled oscillators), $y_0(t)$. We denote this period as $T_0$.
We can derive the Floquet multipliers, $\rho_{1}$ and $\rho_{2}$, by solving the equations above with two different sets of initial conditions \cite{Kovacic2018}. 
The first initial condition is: 

\begin{align}\label{eq:1stic}
\left( 
\begin{array}{c}
\zeta_{k}^{\left( 1\right) }\left( t=0\right)  \\ 
\dot{\zeta}_{k}^{\left( 1\right) }\left( t=0\right) 
\end{array}%
\right) =\left( 
\begin{array}{c}
1 \\ 
0%
\end{array}%
\right) ,
\end{align}

and the second initial condition is:

\begin{align}\label{eq:2ndic}
\left( 
\begin{array}{c}
\zeta_{k}^{\left( 2\right) }\left( t=0\right)  \\ 
\dot{\zeta}_{k}'^{\left( 2\right) }\left( t=0\right) 
\end{array}%
\right) =\left( 
\begin{array}{c}
0 \\ 
1%
\end{array}%
\right) .
\end{align}

The solutions are used to form a matrix 

\begin{align}
B=\left( 
\begin{array}{cc}
\zeta_{k}^{\left( 1\right) }\left( T_0\right)  & \zeta_{k}^{\left( 2\right)
}\left( T_0\right)  \\ 
\dot{\zeta}_{k}^{\left( 1\right) }\left( T_0\right)  & \dot{\zeta}_{k}^{\left(
2\right) }\left( T_0\right) 
\end{array}%
\right) ,
\end{align}

such that the sum of the Floquet multipliers is equal to its trace%

\begin{align}\label{eq:sumfms}
\rho _{1}+\rho _{2}=\zeta_{k}^{\left( 1\right) }\left( T_0\right) +\dot{\zeta}_{k}^{\left( 2\right) }\left( T_0\right) .
\end{align}

In addition, the product of the Floquet multiplies satisfies

\begin{align}\label{eq:prodfms}
\rho _{1}\rho _{2}=\displaystyle\exp \left( \int\limits_{0}^{T_0}tr(A\left( s)\right)
ds\right) =e^{-\gamma T_0}.
\end{align}

These two equations (Eqs. \ref{eq:sumfms} and \ref{eq:prodfms}) provide the Floquet multipliers that determine the
stability of a chain of coupled oscillators to a perturbation with wavenumber $k$. It is important to note that due to the discreteness of the system, the number of resolved wavenumbers (and hence the corresponding resolution in k-space) is set by the number of coupled oscillators in the system. 

\begin{figure}[!h]
\includegraphics[width=\linewidth]{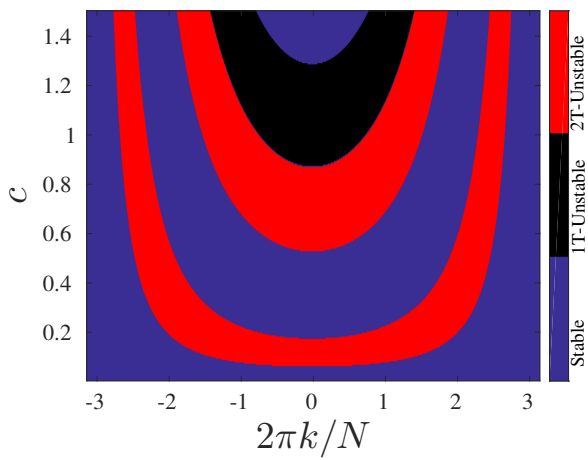}
\caption{\label{fig:1T2Tstab} The stability of the 1T and 2T solutions for $f=13.3$, where both solutions are stable in the absence of coupling. The blue denotes stability, the black denotes instability of the 1T solution, and the red denotes instability of the 2T solution to spatially periodic perturbations with wavenumber $k$ and coupling coefficient, $c$.}
\end{figure}

We derived the Floquet multipliers, for a wide range of wavenumbers, by numerically integrating Eq. \eqref{eq:matpertfdyn} starting with the initial conditions of Eqs. \eqref{eq:1stic} or \eqref{eq:2ndic}. Then we solved Eqs. \eqref{eq:sumfms} and \eqref{eq:prodfms}. If the absolute value of the largest Floquet multiplier is larger than $1$, it implies instability of the uniform solution to spatially periodic perturbation with the corresponding wavenumber. 
\begin{figure}[!h]
\includegraphics[width=\linewidth]{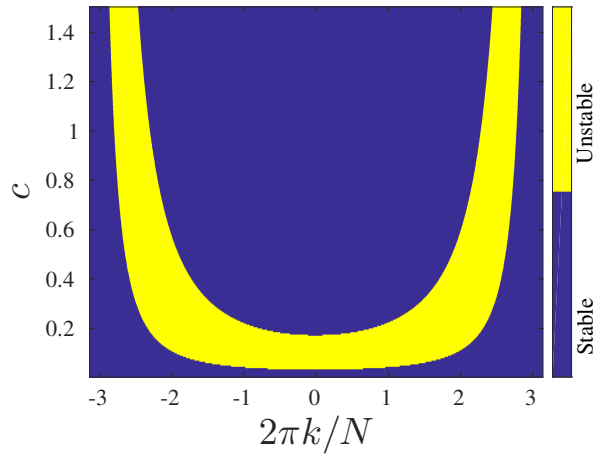}
\caption{\label{fig:1Tstab} The stability of the 1T solution for $f=8.0$, where the 1T is the only stable solution. The blue/yellow colors denote stability/instability to spatially periodic perturbations with wavenumber $k$ and coupling coefficient, $c$.}
\end{figure}
\begin{figure}[!h]
\includegraphics[width=\linewidth]{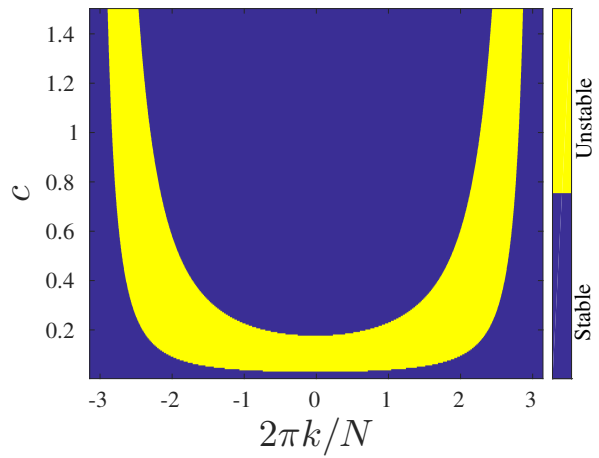}
\caption{\label{fig:2Tstab} The stability of the 2T solution for $f=9.0$, where the 2T is the only stable solution. The blue/yellow colors denote stability/instability to spatially periodic perturbations with wavenumber $k$ and coupling coefficient, $c$.}
\end{figure}
In Figs. \ref{fig:1Tstab} and \ref{fig:2Tstab}, we present the stability diagram as a function of the coupling strength, $c$, and the perturbation wavenumber, $k$, for the $1T$ and $2T$ solutions, respectively. In Fig. \ref{fig:1Tstab}, we used $f=8.0$ for which the only stable solution has the same period as the driving force ($T_0\equiv2\pi/\omega_0$, see Fig. \ref{fig:L2Bif}). In Fig. \ref{fig:2Tstab}, we used $f=9.0$ for which the only stable solution has a period that is twice the period of the driving force ($T_0=4\pi/\omega_0$, see Fig. \ref{fig:L2Bif}). In both figures, it is apparent that for weak coupling, the uniform solution is stable under any perturbation. Above a certain value of the coupling strength, there is wide range of $k$ values for which the uniform solution is unstable. For stronger coupling, only $k$ values within a narrower range destabilize the uniform solution. The symmetry between positive and negative values of $k$ is inherent to the nature of the perturbation and the homogeneous unperturbed model. One can see that the stability diagrams for the $1T$ and $2T$ solutions are similar for driving force amplitudes such that there is only one stable solution (the $1T$ or the $2T$). We used $N=2001$ (which allowed a resolution of $2\pi/2001$ in the resolved wavenumbers), and the coupling strength was modified in steps of $0.05$ in the range of $0$\textendash$1.5$.
In Fig. \ref{fig:1T2Tstab},  we present the stability diagram for a driving force amplitude within the bistability region, $f=13.3$ (see Fig. \ref{fig:L2Bif}). We used the same range and resolution for the wavenumber and the coupling strength as those that were used to generate Figs. \ref{fig:1Tstab} and \ref{fig:2Tstab}. One can easily see that the stability of the uniform $1T$ and the stability of the uniform $2T$ solutions are very different. The uniform $1T$ solution is stable for much larger values of the coupling strength, $c$, compared with the uniform $2T$ solution. One can also notice that for the $1T$ solution, the stability diagram exhibits an additional U-shaped region of instability that did not appear for the cases of a single stable uniform solution.
One can also notice that within the bistability regime, $f=13.3$, the 1T solution is stable under coupling strengths that are much greater than the coupling strength for which it becomes unstable if it is the only stable solution, $f=8.0$ (see Fig. \ref{fig:1Tstab}).

The linear stability analysis is not sufficient because the nonlinear dynamics enables the transfer of energy and momentum between different spatial modes.

\section{Dynamics beyond the initial perturbation: Final states}

\begin{figure}[!h]
\includegraphics[width=\linewidth]{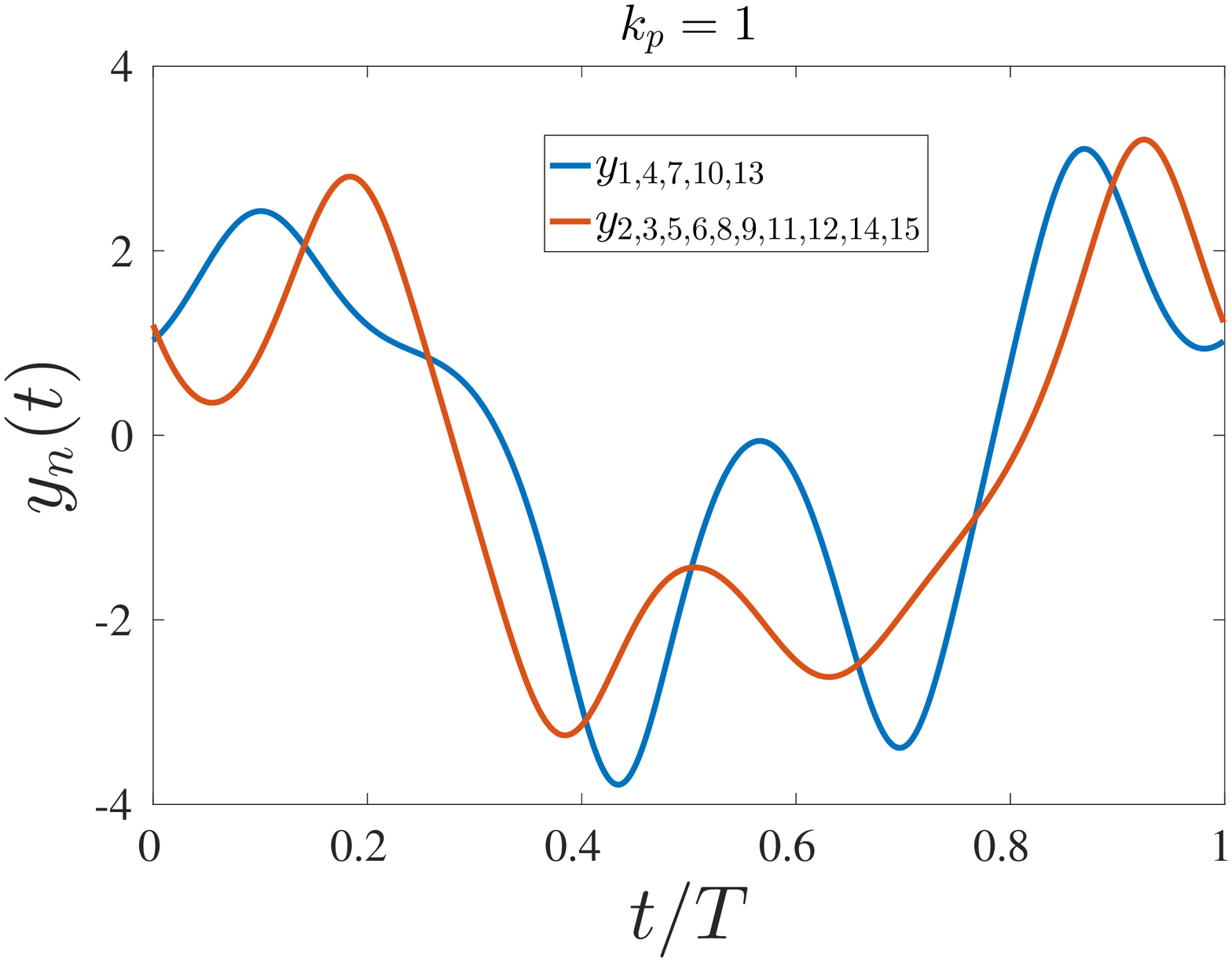}
\caption{\label{fig:1Tk1FS} The solution for 15 coupled oscillators in the final state after the perturbation (1000 periods of the driving force after the perturbation was applied). As can be seen, not all the oscillators are synchronized; namely, the final state is not uniform due to the growth of an unstable mode.}
\end{figure}

\begin{figure}[!h]
\includegraphics[width=\linewidth]{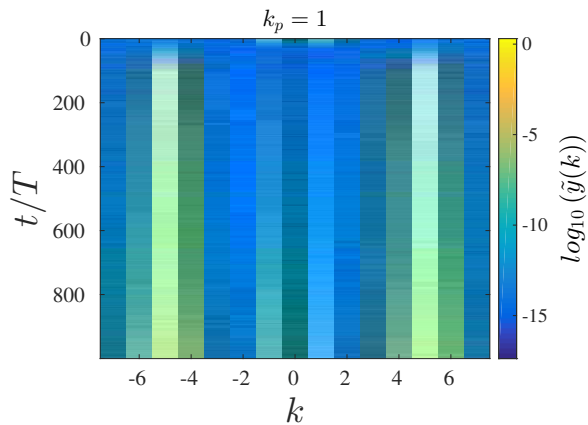}
\caption{\label{fig:1Tk1DFT} The spatial DFT amplitudes for a chain of $N=15$ coupled oscillators with $c=1.5$.
The initial condition is the uniform $1T$ solution, which was perturbed by a spatially non-uniform perturbation proportional to $cos\left(2\pi k_p/N\right)$ with $k_p=1$.}
\end{figure}

The initial dynamics of perturbations, as was described above, does not necessarily represent the final states of the system after a perturbation was applied.
In order to begin investigating the final stable states of the system, we considered the $1T$ uniform solution, with $f=13.3$, for which we calculated the stability (see Fig. \ref{fig:1T2Tstab}) and with a coupling strength given by $c=1.5$. A spatially non-uniform perturbation, with a wavenumber corresponding to $k_p=1$, was applied to the system for an instant, and the system evolved without additional perturbations. For a system of 15 coupled oscillators, the mode $k=1$ is stable according to the Floquet analysis for the given system parameters; in fact, the only unstable mode for these parameters is $k=5$. 
In Fig. \ref{fig:1Tk1FS}, we show the steady state solution for a chain of 15 coupled oscillators for a duration of one period of the driving force (which corresponds to one period of the 1T solution). As can be seen, there are two distinct solutions. Oscillators 1, 4, 7, 10, and 13 are all in phase with each other, and oscillators 2, 3, 5, 6, 8, 9, 11, 12, 14, and 15 are also in phase with each other but show a different solution than the first group listed above. 
In order to gain more insights into the dynamics following the perturbation, we present in Fig. \ref{fig:1Tk1DFT} the spatial discrete Fourier transform (DFT) of the chain of $N=15$ oscillators for different times, following a perturbation with a wavenumber corresponding to $k_p=1$. 
The DFT was calculated at the end of each cycle of the driving force for 1000 cycles, starting at the time at which the perturbation was applied. As can be seen, the initial excitation of the mode $k=1$ led to excitation of the unstable mode $k=5$, and the final state has more than one spatial mode with a non-zero amplitude.

This result suggests that the linear stability analysis is not sufficient for studying the final states of the system because the nonlinear dynamics enables the transfer of energy and momentum between different spatial modes. 

Following spatially non-uniform perturbations, the dynamics is not fully described by the linearized equation, which is used to calculate the Floquet multipliers. The nonlinear interactions result in the excitation of modes other than the initially excited mode and, thus, can lead to the instability of the uniform state even if the perturbed mode decays. To see this, one can derive the full dynamics of the perturbation.
The perturbation is defined as:

\begin{eqnarray}
y_{m}(t) &=&y_{0}(t)+u_{m}(t).
\end{eqnarray}%

Neglecting the coupling between the oscillators, we find for the dynamics of the perturbation:

\begin{align}
\frac{d^{2}u_{m}}{dt^{2}}&=-\gamma \frac{du_{m}}{dt}+(2by_{0}-3y_{0}^{2}-1)u_{m}\\ \nonumber
&+\left( b-3y_{0}\right) u_{m}^{2}-u_{m}^{3}.
\end{align}%

Expanding $u_m$ in a Fourier series%

\begin{align}
    u_{m}=\sum_{k=-M}^{M}\eta_{k}e^{ik\frac{2\pi }{N}m},
\end{align}

where, without a loss of generality, we defined $M\equiv(N-1)/2$ and assumed that $N$ is odd. 
The dynamics can be written as

\begin{align}
\frac{d^{2}\eta_{k}}{dt^{2}}=&-\gamma \frac{d\eta_{k}}{dt}+\left(
2by_{0}-3y_{0}^{2}-1\right) \eta_{k}\nonumber\\
&+\left( b-3y_{0}\right) \sum_{k^{\prime
}=-M}^{M}\eta_{k-k^{\prime }}\eta_{k^{\prime }}\nonumber\\
&-\sum_{k^{\prime
}=-M}^{M}\sum_{k^{\prime \prime }=-M}^{M}\eta_{k-k^{\prime }}\eta_{k^{\prime }-k^{\prime \prime }}\eta_{k^{\prime \prime }}.
\end{align}

When the initial perturbation corresponds to a cosine or sine with a single wavenumber $k_{p}$, the only non-zero amplitudes, at the initial time, is $\eta_{\pm k_{p}}.$ Consequently, the only wavenumbers for which the RHS is not zero are:

\begin{align}
\frac{d^{2}\eta_{\pm k_{p}}}{dt^{2}}&=-\gamma \frac{d\eta_{\pm k_{p}}}{dt}+\left(
2by_{0}-3y_{0}^{2}-1-3 \eta_{-k_{p}}\eta_{k_{p}}\right) \eta_{\pm k_{p}}\nonumber\\
    \frac{d^{2}\eta_{\pm 2k_{p}}}{dt^{2}}&=\left( b-3y_{0}\right) \eta_{\pm k_{p}}^2,\nonumber \\
\frac{d^{2}\eta_{\pm3k_{p}}}{dt^{2}}&=-\eta_{\pm k_{p}}^3.
\end{align}

Obviously, the nonlinear dynamics results in excitation of the modes with wavenumbers equal to twice and three times that of the perturbed mode.
The number of spatial modes is set by the number of oscillators in the chain, $N$. Therefore, for a small number of oscillators, it is likely that the final state of the system may depend on the wavenumber of the perturbed mode (because the nonlinear interaction is limited to a small number of interacting modes). On the other hand, for a large number of oscillators, one may expect that any initial perturbation will be able to expand through the nonlinear interactions and destabilize the uniform state if there are unstable modes (and obviously any initial perturbation would decay if all the modes are linearly stable).

\begin{figure}[!h]
\includegraphics[width=\linewidth]{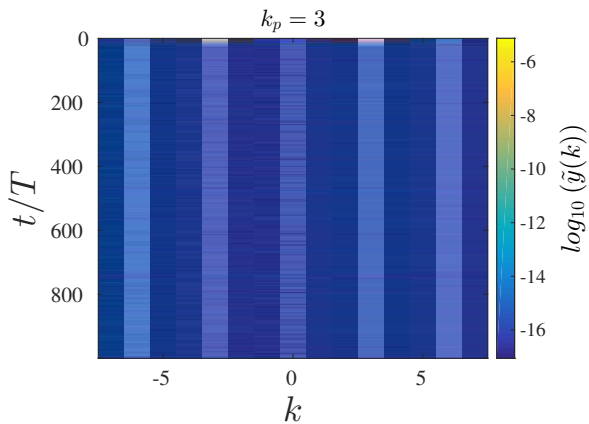}
\caption{\label{fig:1Tk3DFT} The spatial DFT amplitudes for a chain of $N=15$ coupled oscillators with $c=1.5$.
The initial condition is the uniform $1T$ solution, which was perturbed by a spatially nonuniform perturbation with mode $k_p=3$.}
\end{figure}

To illustrate the different responses of uniform states to spatially non-uniform perturbations, we considered several other scenarios. For clarity, we use the same number of coupled oscillators.
Fig. \ref{fig:1Tk3DFT} shows similar information to that shown in Fig. \ref{fig:1Tk1DFT} but for an initial perturbation corresponding to the spatial mode $k_p=3$. In this case, the initially perturbed mode excites only the modes that are integer multiplications of it; specifically, only modes $\pm3$, and $\pm6$ are excited, and these modes are all stable. Therefore, the initial non-uniform perturbation decays, and the system converges to the initial uniform solution.

In Fig. \ref{fig:1Tk1FS}, we show the solution for the 15 oscillators for a duration of one period of the driving force (which corresponds to one period of the solution), 1000 periods after the perturbation was applied. As can be seen, all the oscillators are synchronized, and the chain is spatially uniform.

\begin{figure}[!h]
\includegraphics[width=\linewidth]{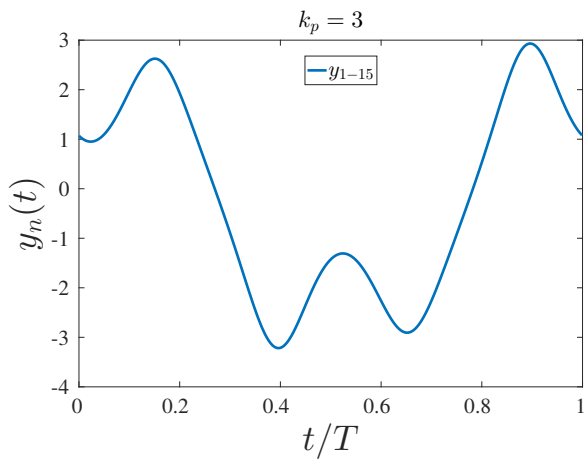}
\caption{\label{fig:1Tk3FS} The solution for the oscillators in the final state after the perturbation. As can be seen, all the oscillators are in-phase; namely, the final state is uniform.}
\end{figure}
\begin{figure}[!h]
\includegraphics[width=\linewidth]{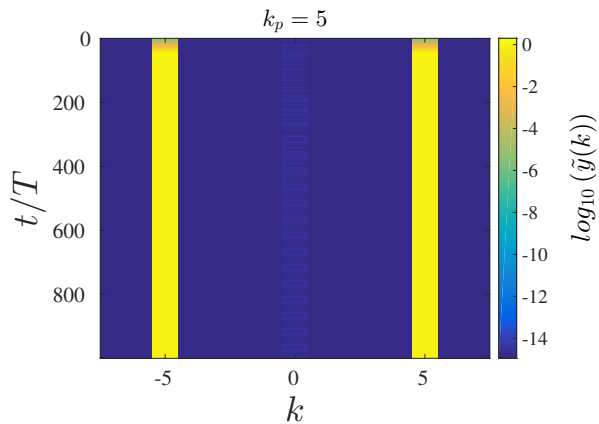}
\caption{\label{fig:1Tk5DFT} The spatial DFT amplitudes for a chain of $N=15$ coupled oscillators with $c=1.5$.
The initial condition is the uniform 1T solution, which was perturbed by a spatially nonuniform perturbation with mode $k_p=5$.}
\end{figure}

In Fig. \ref{fig:1Tk5DFT}, we show the dynamics of the DFT for the case of a perturbation characterized by the wavenumber $k_p=5$. In this case, there is no excitation of other wavenumbers, and the final state is also characterized by the perturbed wavenumber.

The final state of the oscillators in this case is very similar to the one shown in Fig. \ref{fig:1Tk1FS}, which is also dominated by the same wavenumber.

\begin{figure}[!h]
\includegraphics[width=\linewidth]{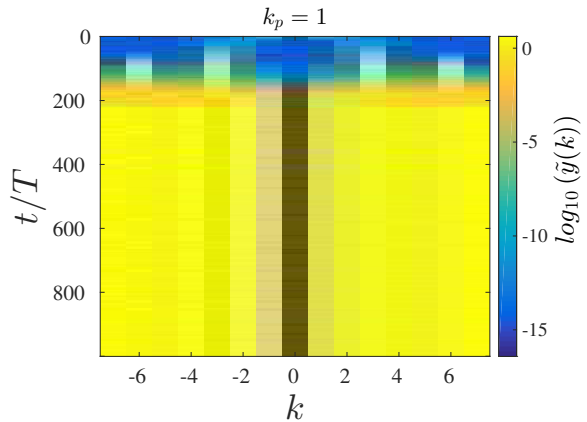}
\caption{\label{fig:2Tk1DFT} The spatial DFT amplitudes for a chain of $N=15$ coupled oscillators with $c=0.3$.
The initial condition is the uniform 2T solution that was perturbed by a spatially non-uniform perturbation with mode $k_p=1$.}
\end{figure}

For the $2T$ uniform state, the responses to spatially non-uniform perturbations are somewhat similar to those found above (for the uniform $1T$ state), but there is one additional type of final state which is different from those shown above. 
In Fig. \ref{fig:2Tk1DFT}, we present the spatial DFT of a chain with $N=15$ oscillators for different times. The DFT was calculated at the end of every two cycles of the driving force for 2000 cycles, starting at the time at which the perturbation was applied. Despite the non-uniform spatial perturbation, all the oscillators remained with a period of $2T$. The driving force is the same as the one used above, $f=13.3$, and the coupling strength is set by $c=0.3$. As can be seen, the initial excitation of the mode $k_p=1$ leads to excitation of the unstable mode $k=3$, which, in turn, excites another unstable mode $k=6$, and the system ends in a non-uniform spatial state that is a mixture of all the possible spatial modes.

\begin{figure}[!h]
\includegraphics[width=\linewidth]{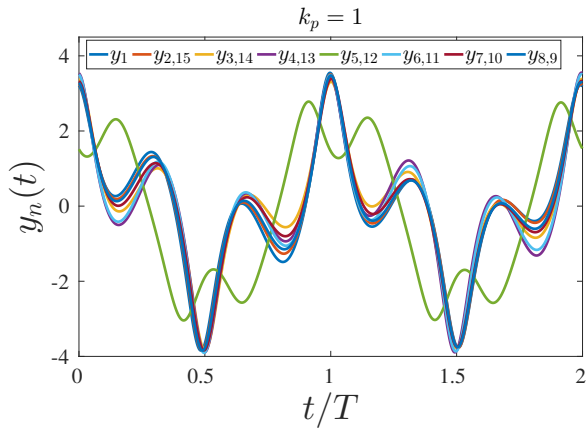}
\caption{\label{fig:2Tk1FS} The solution for the oscillators in the final state after the perturbation. As can be seen, the oscillators are not synchronized, and the final state is not uniform.}
\end{figure}

In Fig. \ref{fig:2Tk1FS}, we show the solutions for the 15 oscillators for a duration of two periods of the driving force (which correspond to one period of the solutions). As can be seen, there are eight different solutions. 

\section{Discussion}
The emergence of patterns due to finite wavenumber instabilities is well known and has been extensively studied in various systems including coupled nonlinear oscillators. However, the effects of multistability on the emergence of spatial patterns is not well understood and has received much less attention. Here, we used the coupled Helmholz-Duffing oscillators to investigate stability to spatially non-uniform perturbation within a parameter regime exhibiting bi-stability for uncoupled oscillators. 

We suggested a numerical method for the calculation of the Floquet multipliers for any uniform solution, even within multistability regions. We found that the uniform state of oscillations in the same frequency as the driving force (1T) is more stable than the uniform state of oscillations with half the frequency of the driving force (2T). The uniform 1T state only loses stability when the coupling is very strong, while the 2T loses stability for a much weaker coupling, suggesting that the existence of an additional stable state (2T) increaes the stability of the 1T uniform state. For both the 1T and 2T states, we found that in the limit of zero coupling, the system is stable, and as the coupling strength increases, multiple modes (wavenumbers) become unstable. When further increasing the coupling strength, many modes regain stability, and a smaller range of modes remains unstable (see Figures \ref{fig:1Tstab}, \ref{fig:2Tstab} and \ref{fig:1T2Tstab}).

In previous studies, the stability of systems with no multistability was studied but only for specific values of the coupling coefficient \cite{English2012,Xu2014,Palmero2016}. Therefore, the system was found to become unstable when the number of coupled oscillators exceeds a certain number. The stability analysis presented here is valid for any number of coupled oscillators. Due to the discrete number of oscillators, the resolved wavenumbers are well defined. Therefore, if the unstable wavenumbers are not resolved for a specific number of coupled oscillators, the system will remain stable.

The final stable states following a spatially non-uniform perturbation seem to have different characteristics depending on the nonlinear energy transfer mechanisms and the resolved wavenumbers. We found spatially periodic states (e.g., Figs. \ref{fig:1Tk1FS} and \ref{fig:1Tk1DFT}), localized states, and states that seem to be spatially disordered (e.g., Figs. \ref{fig:2Tk1DFT} and \ref{fig:2Tk1FS}).

The lack of synchronization, commonly as a result of a phase difference between the different oscillators, may also stem from differences in the frequency of the oscillators when states with different frequencies are stable. However, we found that single mode perturbations do not lead to the formation of double period breather states \cite{Bel2018double}. This remained true even when we varied the model parameters (the friction, the asymmetry parameter, and the frequency of the driving force). Therefore, the conditions for the existence and stability of double period breathers seem to be unrelated to the stability of synchronized states. Different approaches are required to improve our understating of these double period breathers.
\section*{Acknowledgement}
The research of BSA, K{\O}R, and ARB reported in this publication was supported by the National Institutes of Health under award number R01MH116281 

\bibliographystyle{elsarticle-num} 
\bibliography{breathers4}








\end{document}